\documentclass[10pt]{article}
\usepackage[OE]{express}

\begin{document}
\title{Endoscopic imaging of quantum gases through a fiber bundle}

\author{Daniel Benedicto-Orenes, Anna Kowalczyk, Kai Bongs, and Giovanni Barontini\authormark{*}}

\address{Midlands Ultracold Atom Research Centre, School Of Physics and Astronomy, University of Birmingham, Edgbaston, Birmingham, B15 2TT, United Kingdom}

\email{\authormark{*}g.barontini@bham.ac.uk} 



\begin{abstract}
We use a coherent fiber bundle to demonstrate the endoscopic absorption imaging of quantum gases. We show that the fiber bundle introduces spurious noise in the picture mainly due to the strong core-to-core coupling. By direct comparison with free-space pictures, we observe that there is a maximum column density that can be reliably measured using our fiber bundle, and we derive a simple criterion to estimate it. We demonstrate that taking care of not exceeding such maximum, we can retrieve exact quantitative information about the atomic system, making this technique appealing for systems requiring isolation form the environment.          
\end{abstract}

\ocis{(020.1475) Bose-Einstein condensates; (110.2350) Fiber optics imaging; (170.2150) Endoscopic imaging.} 


\section{Introduction}

Cold atoms systems are at the core of the emerging quantum technologies and represent an invaluable resource for the exploration of quantum phenomena. On the one hand they enable the development of increasingly precise sensors and measurement devices, on the other hand they can be manipulated with great control, allowing the engineering of complex hamiltonians for quantum simulation. Often, to achieve such exquisite level of control and precision, cold atoms systems need to be isolated from the environment. Magnetically sensitive experiments, such as atom magnetometers \cite{kornack_low-noise_2007,kk}, atom clocks \cite{moric_magnetic_2014}, atom interferometers \cite{kovachy_quantum_2015} and quantum simulators \cite{bloch_quantum_2012}, need to be accurately shielded from external magnetic fields using one or more layers of $\mu$-metal. Experiments that require cryogenic temperatures \cite{Haroche, bernon_manipulation_2013} need instead to be performed inside cryostats. Thermal isolation is also desirable to increase the performance of atomic clocks \cite{beloy_atomic_2014}.

Within the experiments aiming at realizing quantum simulations, there are those that require extreme isolation from the environment to explore the effects of dipolar interactions in quantum gases \cite{Vedmedenko, Ueda}. Unless one employs atomic species with permanent magnetic dipole moment like Dy \cite{Dy} or Er \cite{Er}, to investigate such effects it is necessary to ensure that the interaction energy of the atomic dipole moments is not washed out by the Zeeman coupling to any residual magnetic fields. For example, since in $^{87}$Rb the dipole-dipole energy is only $\simeq h \times 1$Hz, with $h$ the Planck constant, and the Zeeman splitting goes approximately as $\simeq h\times7\times10^9$ Hz$/$T, it follows that the external magnetic field should be below $10^{-9}$T for dipolar effects to become relevant. Our experiment has been built to explore such extreme regime and has been designed to be completely non magnetic and to accommodate 5 layers of $\mu$-metal magnetic shield, in combination with active field compensations.

In our experiment, as in all those that require a strong shielding, retrieving information from the system inevitably leads to an unwanted coupling with the environment. Particularly disruptive is the use of the widely employed absorption imaging, since it requires high numerical aperture optics and sophisticated CCD cameras. These latter are not compatible with a shielded environment so they need to be accommodated outside the shield. Then, to allow the image of the atoms to reach the camera, one or more holes must be cut in the shield resulting in a detrimental loss of shielding factor \cite{donley_demonstration_2007}. 

In this work, we follow a different approach and we demonstrate the use of a fibre bundle to perform endoscopic absorption imaging of quantum degenerate gases. Fibre bundles are fibre optic devices composed of thousands of standard optical fibres which are packed together. They are widely used in biological applications for fluorescence imaging techniques \cite{Rouse, Shin, fiber-optic_2005} and in medical endoscopy for optical coherence tomography and multiphoton microscopy \cite{confocal_1,confocal_2,two_photon,confocal_3}. If the spatial ordering of the fibres is preserved on both ends, the bundle is regarded as coherent and can be used to transfer an image from one end to the other. Fibre bundles have a diameter of only a few mm and are flexible, making them ideal candidates to transport the absorption images through small holes in the shield, therefore causing only minor disruption in the shielding. By accounting for the spurious effects introduced by the fibre bundle, we show that quantitative information can be retrieved from the absorption pictures, making our technique extremely appealing for experiments requiring high shielding factors.

\section{The experimental apparatus}

Our experiment aims at studying the magnetism of $^{87}$Rb in extremely low magnetic fields (~10$^{-9}$ T), therefore the apparatus is entirely made of non-magnetic materials and has been designed to be surrounded by a five-layer $\mu$-metal shield. The atoms are initially pre-cooled in a 2D-MOT, where the chamber is made out of a single titanium block. Using a weak resonant beam with the D2 transition of $^{87}$Rb ($\simeq$780nm) propagating along the axis of the 2D MOT, we push the atoms through a differential pumping stage towards a second, 50 cm long, all-glass chamber. At the end of this latter, we collect $~10^8$ atoms in 3 s in a 3D MOT. The typical temperature of our MOT is $\simeq$300 $\mu$K. The magnetic field gradient is provided by a small pair of anti-Helmholtz coils mounted on a plastic support. At the end of the MOT stage, our sequence includes a 80 ms dark MOT stage, during which we reduce the repumper power from its maximum value to $1.5\%$ of the total power while ramping the detuning of the cooling light towards the red by $\simeq 80$ MHz ($\simeq12{\Gamma}$). The temperature of the atoms at the end of the dark MOT stage is $\simeq$  75$\mu$K, additionally all the atoms result pumped in the $|F=1\rangle$ ground state manifold. During the whole duration of the MOT and dark MOT, a 16 W beam at 1070 nm, focussed in the centre of the atomic cloud with a waist of 33 $\mu$m, is kept on. The beam is produced by a multimode, unpolarized YAG fibre laser (IPG YLR-20) and its power is controlled by a free space acousto optic modulator. At the end of the dark MOT we switch off the magnetic field gradient and $\simeq 10^7$ atoms are loaded into the 1070 nm dipole trap with a typical temperature of 150 $\mu$K, set by the depth of the trap. We then start the evaporation in the dipole trap by ramping down the power of the beam in a linear piecewise ramp. In the first stage, we decrease the power from 100$\%$ to 17$\%$ in 260 ms. In a second stage, we ramp it down from 17$\%$ to 1.4$\%$ in a total time of 5.9 s. A second beam, with a waist of  36$\mu$m and aligned to cross the 1070 nm beam at an angle of 80 degrees, is switched on 1.26 s after the end of the dark MOT stage. This beam, that has a wavelength of 1550 nm, is produced by a DBR diode laser, which is amplified by a fiber amplifier (NUA-1550-PB-0010-B0). It is single mode, linearly polarized and has a power of 4.6 W. It is kept constant for the first 1.5 s at 21$\%$ of its total power. After this time, it is also ramped down first to 11$\%$ during 1.5 s and finally to 1$\%$ in 1.9 s. We end up with a pure spinor Bose-Einstein condensate (BEC) of $\simeq 1.5\times 10^5$ atoms in the F=1 spin manifold. The final trapping frequencies of our crossed dipole trap are $2\pi\times(284, 284, 60)$ Hz, with the weaker confinement along the direction of propagation of the strong 1070 nm beam.

\begin{figure}
	\centering
		\includegraphics[width=0.7\textwidth]{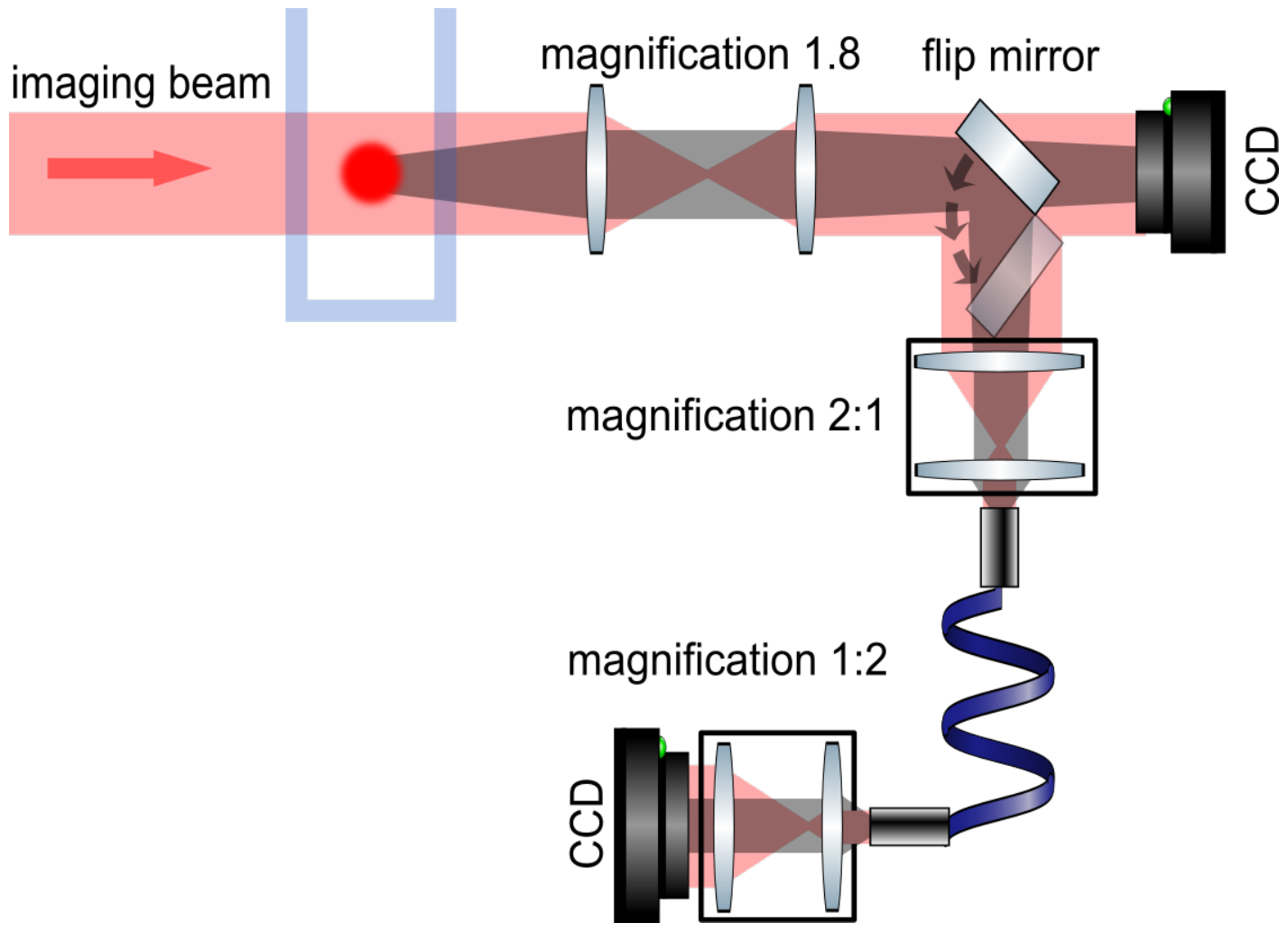}
	\caption{Schematics (not in scale) of the optical setup used. The atomic cloud casts a shadow on the imaging beam that is magnified by a factor of 1.8 by a first telescope. A flipping mirror send the image of the shadow of the atomic cloud either on the CCD or on the fiber bundle setup. In this latter a 2:1 telescope images the focal plane on the input facet of the fiber bundle. A 1:2 telescope then images the output facet on the CCD camera.}
\label{setup}
\end{figure}

\section{Endoscopic absorption imaging}

In quantum gases experiments, absorption imaging is the most commongly used technique to retrieve information about the atomic sample. In our experiment, a collimated beam resonant with the $|F=2\rangle \rightarrow |F'=3\rangle$ transition of the $^{87}$Rb D2 line is sent on the atoms, that absorb the light and leave a "shadow" on the beam that in turn is imaged into a CCD camera (656$\times$492 pixels with size 5.6 $\mu$m) with a magnification factor of 1.8. The amount of light absorbed by the atoms is given by the Beer-Lambert law: $I=I_0\exp[-\int n(z)\sigma dz]$, where $z$ is the direction of propagation of the beam, $I_0$ the intensity of the beam, $n(z)$ the density of the cloud and $\sigma=\sigma_0/(1+4(\Delta/\Gamma)^2+I_0/I_s)$ the absorption cross section, with $\Delta$ the detuning from the atomic transition, $\Gamma$ the width of the transition, $I_s$ the saturation intensity and $\sigma_0=\hbar\omega\Gamma/(2 I_s)$ the resonant cross section, $\omega$ being the frequency of the transition. The quantity $\Omega=\int ndz$ is the column density. Besides the image with the atoms $I_1$, for each repetition of the experiment we take also a picture of the imaging beam without atoms $I_2$ and a picture with no light and no atoms $I_3$. The column density of the atomic cloud is then retrieved pixel by pixel as $\Omega(x,y)= -ln[I(x,y)/I_0(x,y)]= -ln[(I_1(x,y)-I_3(x,y))/(I_2(x,y)-I_3(x,y))]$ \cite{ketterle}.

\begin{figure}
	\centering
		\includegraphics[width=\textwidth]{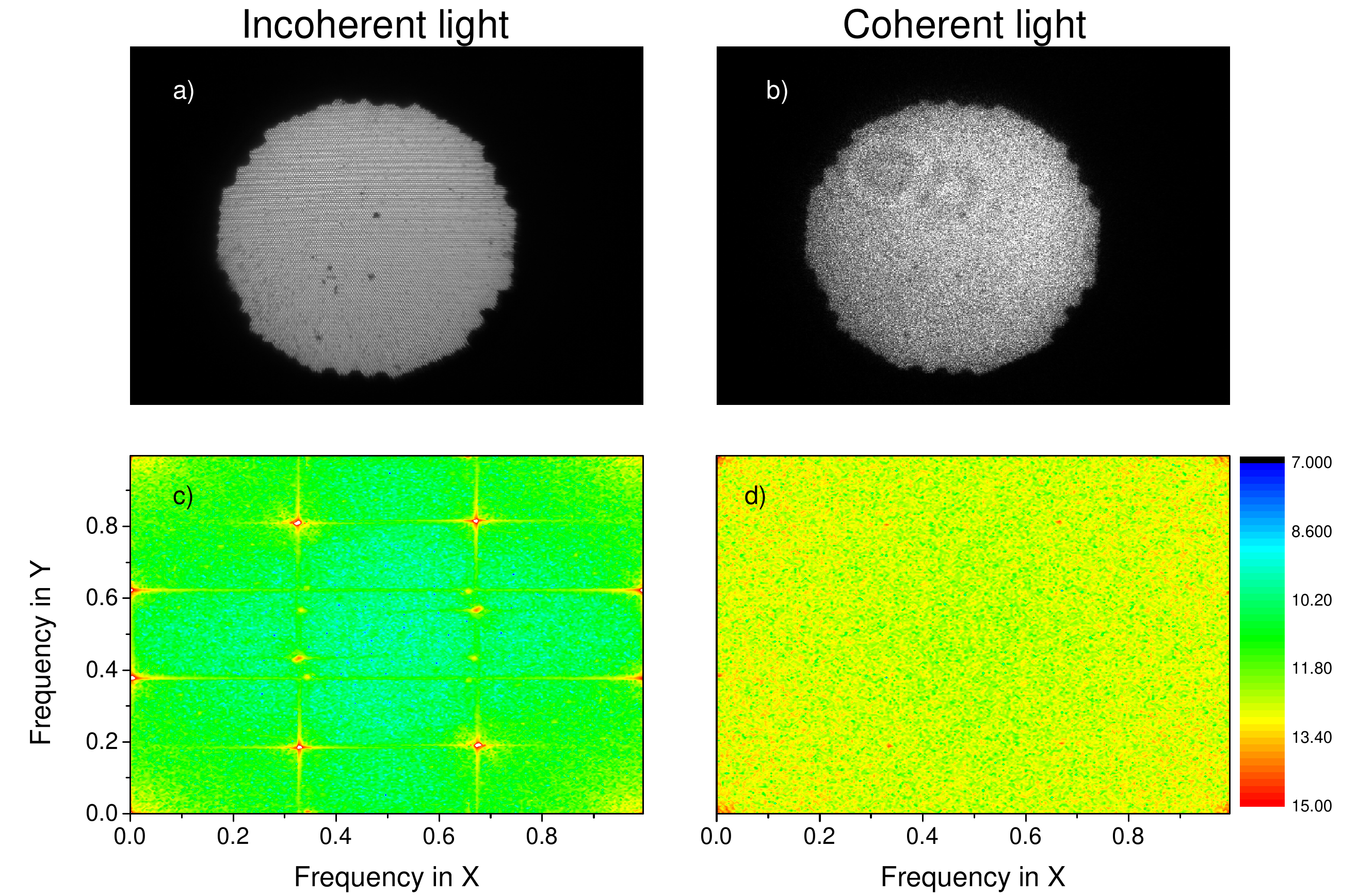}
	\caption{Image of the output facet of the fiber bundle when injected with incoherent (a) and coherent light (b). c) and d) are the corresponding two-dimensional Fourier transforms. The color scale is in dB. While in a) it is possible to appreciate the hexagonal packing of the fiber bundle, in b) only a speckle pattern is visible. Equivalently, in c) clear peaks emerge from the spectrum while in d) no peaks are visible.}
\label{light}
\end{figure}

As shown in Fig. 1, to perform the endoscopic imaging of our ultracold sample, instead of sending the light directly to the CCD camera, we send it to a leached fiber bundle. This latter is a commercially available fiber bundle (Schott 1249311 RLIB CVET,1.00 X 670,8.2M,13.5K,QA 0.80) that is 67 cm long and contains a total of 13500 fibers with 8.2 $\mu$m core diameter. The fibers are coherently packed in an hexagonal pattern so that the input/output facet at each end has 1 mm outer diameter. After magnification, the focal plane of the atoms is imaged on the input facet of the bundle with a 2:1 telescope made of two achromatic lenses (Thorlabs MAP1050100-B). A similar optical set up with a 1:2 magnification is used to image the output facet of the fiber bundle onto the CCD chip. It is formed by two lenses: a high NA (0.5) f = 8 mm aspheric lens (Thorlabs AC240TME-B) and a f = 16 mm lens (Thorlabs AC080-16-B). It is mounted on a cage system to allow for precision alignment. This allows us to properly collect the light coming out the bundle since the typical NA of the fibers is $\approx$ 0.35.  With this setup, the overall magnification for the fiber bundle optical system is also 1.8. To characterize the absorption imaging through the bundle, we use a simple flipping mirror that allows to switch between the two imaging paths and therefore to have the direct comparison of the pictures transmitted through the bundle with the pictures taken in free space.   

By passing through the fiber bundle, pictures can suffer from spurious effects such as multimodal coupling \cite{han_effect_2012}, cross talk between the fiber cores \cite{chen_experimental_2008} and pixelation due to the packing structure of the bundle that can potentially alter the information transmitted. All this, in addition to a non-unitary quality area, also severely limit the transmission efficiency \cite{udovich_spectral_2008}. 

To limit pixelation and have a better image quality we have chosen a bundle with a large number of small core fibers closely packed. The price to pay is a non-negligible core-to-core coupling. We have found that the cross talking between fibers must be accounted especially when working with coherent laser light. 
In Fig. \ref{light}(a) and (b) we report an image of the output facet of the bundle when it is injected with incoherent and coherent light respectively. In the case of incoherent light, the underlying matrix of fibers is clearly visible. This becomes even more clear when performing the 2D Fourier transform of the image, see Fig. \ref{light}(c), where the regular structure of the fibre packing give rise to clear peaks in the power spectrum. When instead we use coherent light, a speckle pattern appears and the regular lattice of fibres is completely washed out. This is due to the fact that the laser light acquires a different phase inside each different fiber. The cross-talking between different fibers inside the bundle creates the interference that generates the speckle pattern. By performing the 2D Fourier transform we confirm the absence of the peaks on the power spectrum, see Fig. \ref{light} (d) and we observe that the noise level is significantly increased. The appearance of a noisy speckle pattern is not \emph{per se} a problem for the absorption imaging since it is stationary and it is eliminated when computing $\Omega$. However we will see in the following that the percolation of light into adjacent fibers severely affects images of objects with high optical density. 

\begin{figure}
	\centering
		\includegraphics[width=0.7\textwidth]{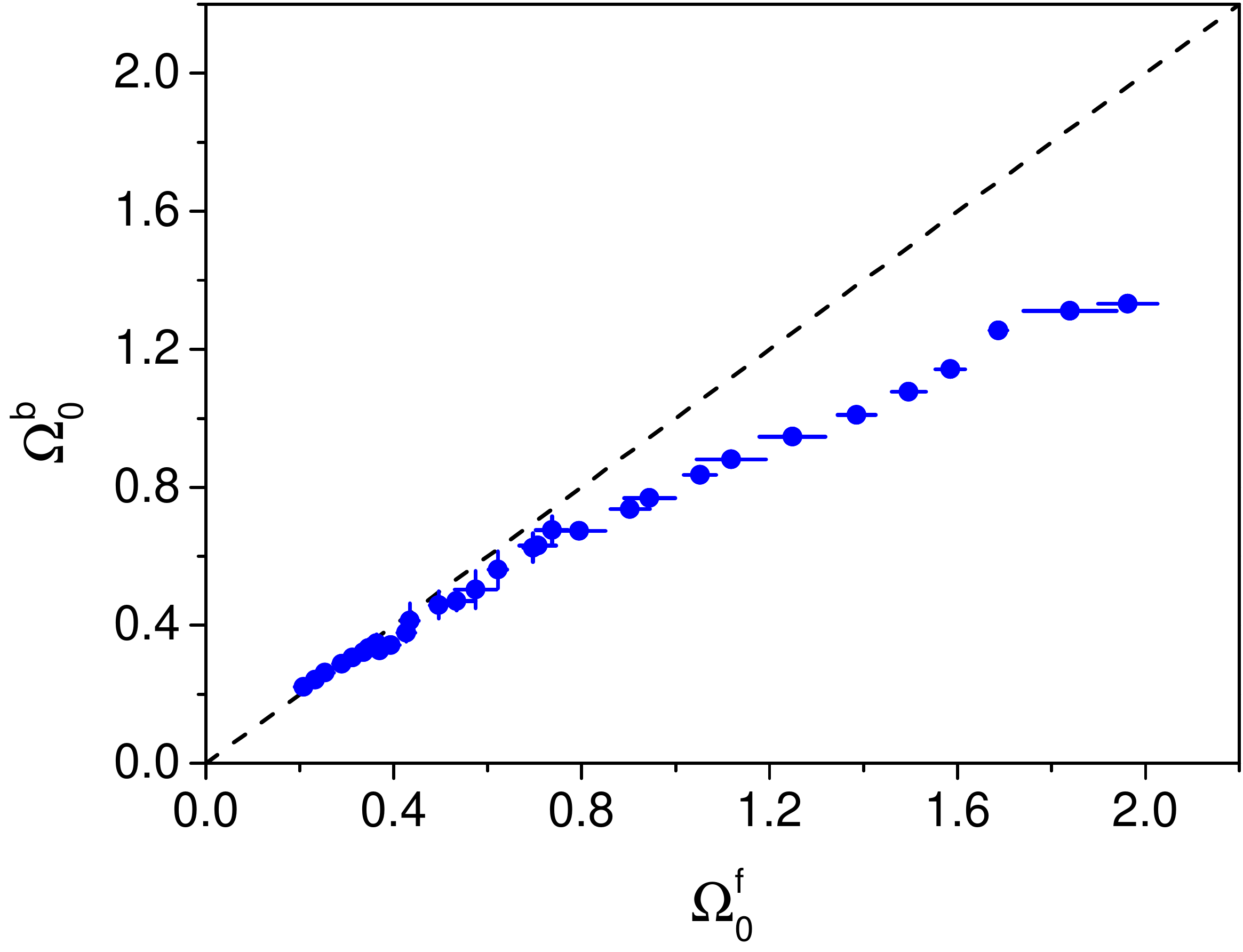}
	\caption{Peak column density measured using the fiber bundle $\Omega_0^b$ as a function of the peak column density measured in free space $\Omega^f_0$ for the same atomic sample. The dashed line is the curve $\Omega^b_0=\Omega^f_0$. The errorbars are one standard deviation statistical errors.}
\label{omega}
\end{figure}

In principle, another major effect to take into account is the low transmission efficiency of the bundle, which acts as an optical attenuator.
We have measured the efficiency of our fiber bundle to be  $\eta\simeq 0.33$\% at 780nm, which is compatible with the typical values quoted by the manufacturer and in other studies \cite{udovich_spectral_2008}. However it is easy to verify that absorption imaging is robust against this effect. Indeed, in the absence of other spurious effects, even in case of very poor efficiency the column density obtained using the fiber bundle is identical to the one in free space: 
$\Omega^b = -ln[(I^b_1-I^b_3)/(I^b_2-I^b_3)]=-ln[\eta(I^f_1-I^f_3)/(\eta(I^f_2-I^f_3))] = \Omega^f$, where the superscripts $b$ and $f$ stand for bundle and free-space respectively. 

To measure the effect of the cross-talk and of the transmission efficiency we take absorption pictures of our atomic clouds at different temperatures and number of atoms. For each set of parameters we vary the intensity of the probe beam from $\simeq$ 0.05  to 0.7 $I_0/I_s$. To make a direct comparison, we take the absorption picture of the cloud with the same parameters, both using the fiber bundle and in free space. We then fit each absorption image $\Omega(x,y)$ using a 2d Gaussian function and extract the peak column density $\Omega_0$, that is directly proportional to the number of atoms in the cloud $N=2\pi\Omega_0\sigma_x\sigma_y/\sigma_0$, where $\sigma_{x,y}$ are the widths of the Gaussian distribution and $\sigma_0$ is the resonant scattering cross section. We have verified that, within our errorbars, $\sigma_{x,y}$ measured through the fiber bundle are identical to those measured in free space, as expected since the overall magnification of the two imaging system is equal.    

\begin{figure}
	\centering
		\includegraphics[width=\textwidth]{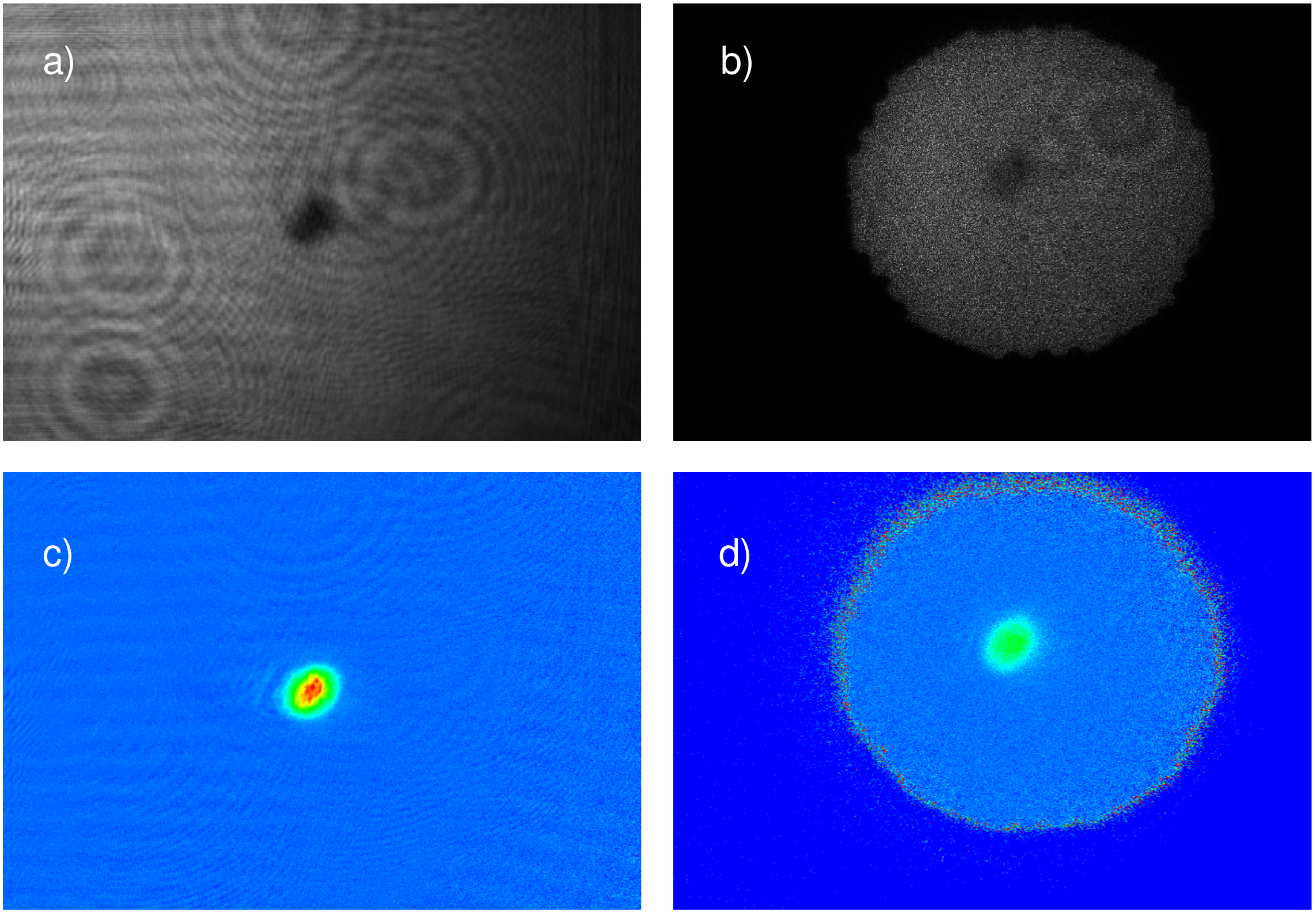}
	\caption{a) and b) Direct picture of the absorption profile of a Bose-Einstein condensate in free space ($I^f_1$) and through the fiber bundle ($I^b_1$) respectively. c) and d) the corresponding absorption pictures $\Omega^f$ and $\Omega^b$. The colour scale is the same in c) and d). Each picture is 656$\times$492 pixels.}
\label{picture}
\end{figure}

In Fig. \ref{omega} we summarize our results reporting $\Omega_0^b$ as a function of $\Omega_0^f$. We observe that, within our errorbars, $\Omega^b\equiv\Omega^f$ for values below $\Omega_M\simeq$ 0.7. We conclude therefore that for small column densities the impact of the fiber bundle in the images has a negligible effect. For values of $\Omega$ above 0.7 however, the peak column density measured with the bundle starts to be lower than the one measured in free space, and the difference increases as $\Omega^f$ increases. Note that we have taken care that no saturation effects were present in the free-space picture and consequently in the pictures taken through the bundle. Indeed this latter transports the light only after it has interacted with the atoms, so it does not play any role in the atom-light interaction. The reason for the discrepancy at higher $\Omega$ can be understood from Fig. \ref{picture}(a) and (b), where we report $I^f_1$ and $I^b_1$ for a dense BEC. In particular, in Fig. \ref{picture} (b) it is possible to observe that even in case of a very dense sample, the atomic shadow is not completely dark, as it is in free space -Fig. \ref{picture} (a). As described above, when passing through the bundle the image acquires a faint speckle pattern created by the core-to-core coupling. This percolates into the fibers that transport the shadow of the atoms preventing this latter to be completely dark on the output facet. This effect is therefore more important in the pictures with high column density samples. The resulting effect can be observed in the absorption pictures in Fig, \ref{picture}(c) and (d) (same colour scale), where the column density measured with the bundle is significantly lower than the one measured in free space.   

Whenever it is possible to make a direct comparison, like in our case, it is easy to correct for the effect of the cross-talking by doing a simple calibration.
For example, one can fit the curve $\Omega\equiv\Omega^f=\Omega^b(\Omega^f)$. However problems arise when this direct comparison cannot be done, as it will be the case once our setup will be enclosed in the five-layer $\mu$-metal shield. In that case the only information available will come from the pictures taken through the bundle. 

A quantitative description of the core-to-core coupling is a challenging task, especially for thousands of fibres and this goes beyond the scope of this work. Here instead we derive a simple criterion to ascertain the maximum reliable column density $\Omega_M$ that can be measured with the fiber bundle, using \emph{only} the information available through the fiber bundle itself. For very dense samples, that in normal conditions would feature a high column density, it is possible to remain below $\Omega_M$ and therefore to obtain correct quantitative information either using a probe beam well above the saturation intensity or increasing the detuning of the probe light $\Delta$. 
If on the one hand remaining below $\Omega_M$ guarantees to extract correct information, on the other it inevitably penalizes the signal-to-noise ratio. We will see in the following that, at least for the parameters of interest for our experiment, the increase in the signal-to-noise ratio is negligible. 

\begin{figure}
	\centering
		\includegraphics[width=\textwidth]{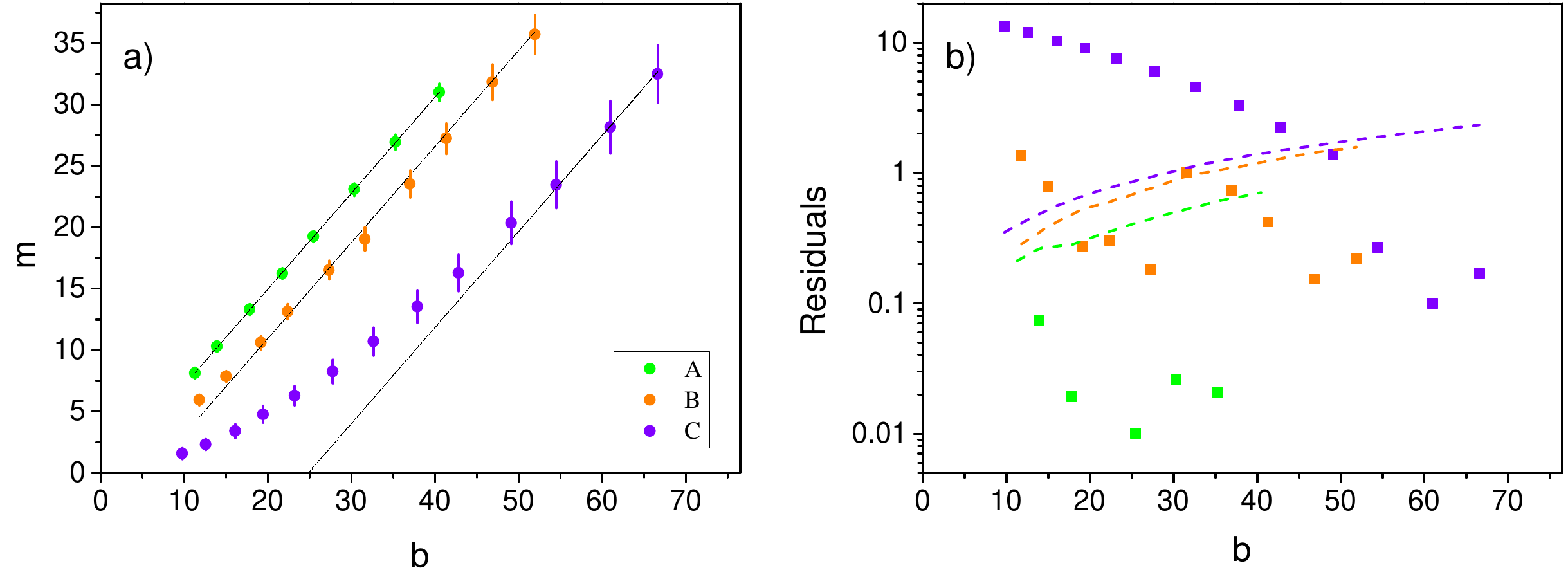}
	\caption{a) Number of counts in the region of maximum absorption $m$ as a function of the average number of counts in the background $b$, evaluated from the direct picture $I_1^b$. All the units are counts per pixel. The three data sets are described in the text. The solid lines are linear fits to the data using the procedure described in the text. The errorbars are one standard deviation statistical errors. b) Absolute value of the residuals of the linear fits reported in a) as a function of $b$. The dashed lines correspond to the errorbars in a).}
\label{linear}
\end{figure}

Our method requires to prepare samples with different densities in order to span a wide range of column densities. In this work the three samples are A=(1.8$\times10^5$, 250 nK); B=(1.2$\times10^5$, 100 nK) and C=(1$\times10^5$, 20 nK), where the first number indicates the number of atoms and the second the temperature of the cloud. Samples A and B are dilute thermal clouds while sample C is a BEC, and they are prepared changing the final point of the evaporation ramp. For each sample we perform the absorption imaging  with different intensities of the probe $I_0$ (in alternative one can scan the detuning $\Delta$). We then perform a Gaussian fit of the shadow cast by the atoms in $I_1^b(x,y)$, retrieving the value of the background light $b$ and the amplitude of the Gaussian $s$ \cite{footnote}. We verify that the value of $b$ coincides with the one obtained from $I_2^b(x,y)$. In the region of maximum absorption  the number of counts in the CCD drops from $b$ to $m=b-s$, and it is easy to verify that $\Omega_0\simeq-ln(m/b)$. In the absence of spurious effects and for $I_0<I_s$, we expect both $s$ and $b$ to increase linearly with the probe intensity $I_0$ and therefore that $m$ is a linear function of $b$. As reported in Fig. \ref{linear}(a), this is the case for the low density sample A. From a linear fit to the data set of the sample A, we obtain the slope $m/b$ that should be common to all data sets not significantly affected by the core-to-core coupling. We then fit the data sets corresponding to samples B and C using a linear function keeping the offset as the only free parameter, as shown in Fig. \ref{linear}(a).
As a criterion to estimate $\Omega_M$, we discard all the points whose absolute value of the residuals is larger than the errorbars,  corresponding to the points above the dashed lines in Fig. \ref{linear}(b). Using this simple method that relies only on the information acquired through the bundle, we find that $\Omega_M\simeq0.7$, in agreement with what we have observed from the direct comparison with the free space pictures. 

\begin{figure}
	\centering
		\includegraphics[width=\textwidth]{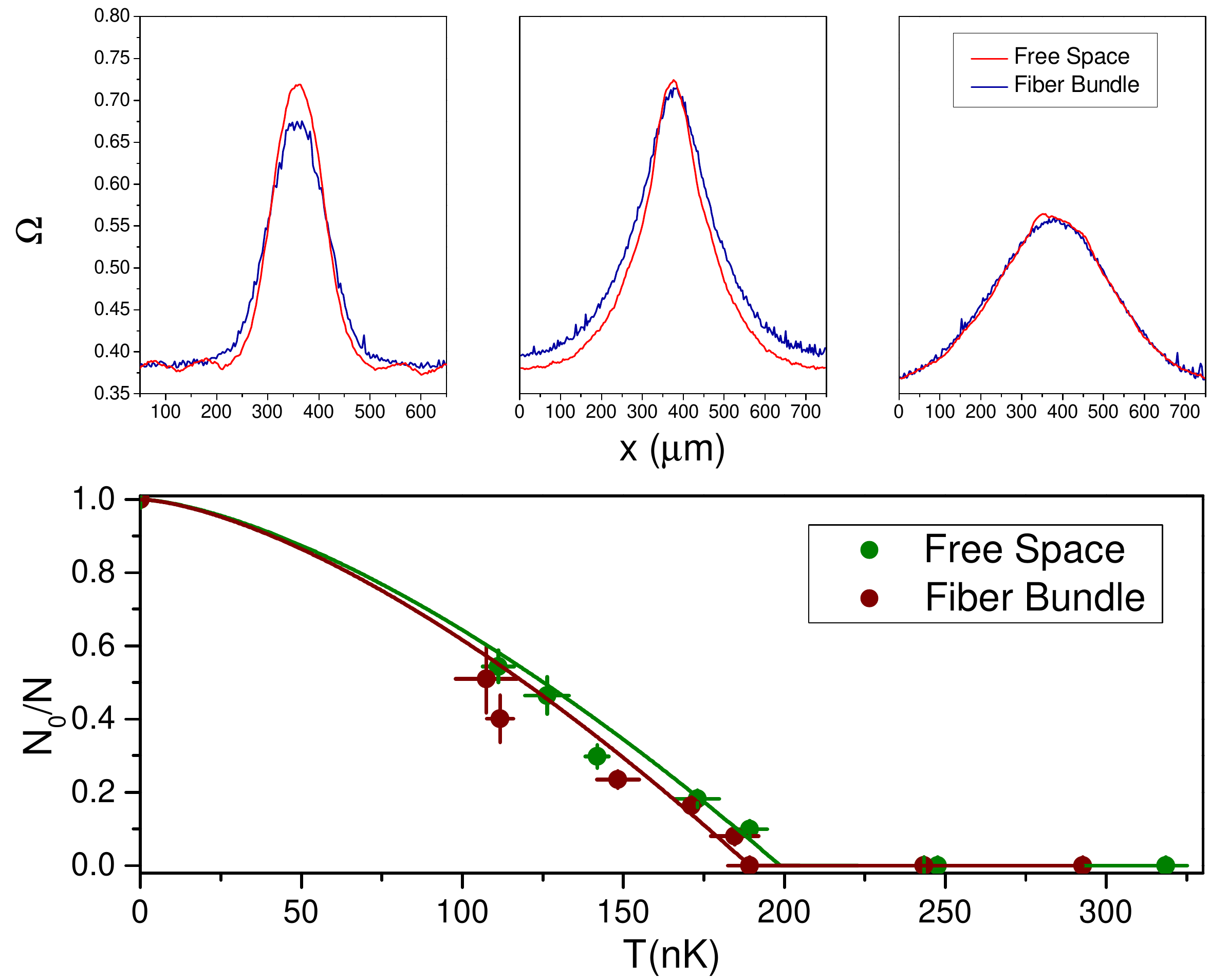}
	\caption{The Bose-Einstein condensate transition. Upper row: integrated density profiles of our atomic sample across the BEC transition measured using the fiber bundle and in free space. From the right to the left: a thermal cloud, a partially condensed cloud and a pure BEC. Lower row: condensate fraction as a function of the temperature. The lines are the fit functions explained in the text.}
\label{transition}
\end{figure}

To give a specific example we have measured the BEC transition using both imaging systems, taking care not to exceed $\Omega_M$ while using the fiber bundle. To this end we have measured the fraction of condensed atoms $N_0$ as a function of  the temperature. This is done fitting $\Omega(x,y)$ with a two-dimensional bimodal distribution made by the sum of a Thomas-Fermi and a Gaussian profile. As can be seen from Fig. \ref{transition}, the results obtained using the fibre bundle are in very good agreement with those taken in free space. For both imaging systems, we are not able to detect any thermal component for condensed fractions above $\simeq60$\%. From the (integrated) density profiles reported in Fig. \ref{transition}, it is possible to appreciate the modest increase in the signal-to-noise ratio using the fiber bundle, that does not significantly affect our measurements. For example, in the leftmost panel of the upper row of Fig. \ref{transition}, the signal-to-noise ratio drops from 168 (free-space) to 104 (fiber bundle). More quantitatively, we compare the critical temperatures $T_c$ measured with the two methods. In three dimensions, the condensate fraction scales as $N_0/N=1-(T/T_c)^3$, with $k_BT_c=0.94\hbar\bar{\omega}N^{1/3}$, where $\bar{\omega}$ is the average trapping frequency and $k_B$ the Boltzmann constant \cite{Stringari}. In an optical dipole trap, the evaporation has two effects: it lowers the temperature of the cloud and reduces the trapping frequencies. Indeed, if $U$ is the depth of the trap (proportional to the trapping laser intensity), we have that $k_BT\simeq\xi U$, with $\xi\simeq1/6$, and $\bar{\omega}=U^{1/2}[(4/mw_0^2)(2/mz_R^2)^{1/2}]^{1/3}$, being $m$ the mass of the atom and $w_0$ and $z_R$ the waist and the Rayleigh length of the trapping beam \cite{grimm}. From this, it follows that for our trap the condensate fraction scales as $N_0/N=1-(T/\tilde{T_c})^{3/2}$, with $\tilde{T_c}=1.88\hbar^2N^{2/3}/[k_B\xi m (w_0^2z_R)^{2/3}]$. By fitting the two curves shown in Fig. \ref{transition} we obtain that $\tilde{T_c}^b=189\pm3$ nK and $\tilde{T_c}^f=198\pm1$ nK, whose difference is within our errorbars $\pm7$ nK, demonstrating the reliability of absorption imaging through the fiber bundle. 

\section{Conclusions}

In this work, we have investigated the use of a fiber bundle to perform endoscopic imaging of quantum gases. We have found that the pictures taken using the bundle are mainly affected by low transmission efficiency and core-to-core light coupling, that creates a noisy speckle pattern. From a direct comparison with free space images, we have found that while performing absorption imaging, the core-to-core coupling strongly affects the images where the column density of the sample is above a certain threshold $\Omega_M$, leading to a systematic underestimation of the number of atoms in the cloud. Below $\Omega_M$, neither the transmission efficiency nor the core-to-core coupling lead to significant systematic effects. We have derived a simple criterion for the estimation of $\Omega_M$ using only the information available through the fiber bundle and demonstrated that, taking care that $\Omega^b<\Omega_M$, the pictures transmitted through the bundle can be used to retrieve exact quantitative information about the atomic system. Our result demonstrate the possibility of using a fiber bundle for minimally invasive imaging of cold atoms experiments requiring a high degree of isolation from the environment.

\section*{Funding}
This work was supported by the Engineering and Physical Sciences Research Council Grants No. Ep/J003875/1 and by the School of Physics and Astronomy at the University of Birmingham.

\section*{Acknowledgments}
We are grateful to D. Paul for his help in the early stage of the experiment and we acknowledge fruitful discussions with the members of the Cold Atoms group at the University of Birmingham.

\section*{Data availability} 
The data presented here are available from the research data management system of the University of Birmingham, accessible online at http://epapers.bham.ac.uk/3024/.


\begin{thebibliography}{99}
\bibitem{footnote}
The correct fitting function for a BEC is a Thomas-Fermi profile. We have verified that within our errorbars, using a Gaussian profile does not introduce any appreciable systematic error in the determination of the parameters here discussed.
\end{thebibliography}
\end{document}